\documentclass[pra,superscriptaddress,showpacs,twocolumn,epsf]{revtex4}
\usepackage{epsfig}
\usepackage{graphicx}

\def \E{{\rm e}}

\begin{document}

\title{Teleportation of qubit states through dissipative channels:\\
Conditions for surpassing the no-cloning limit}

\author{\c{S}ahin Kaya \"Ozdemir$^*$}

\affiliation{SORST Research Team for Interacting Carrier
Electronics, 4-1-8 Honmachi, Kawaguchi, Saitama 331-0012, Japan}
\affiliation{CREST Research Team for Photonic Quantum Information,
4-1-8 Honmachi, Kawaguchi, Saitama 331-0012, Japan}
\affiliation{Graduate School of Engineering Science, Osaka
University, Toyonaka, Osaka 560-8531, Japan}

\author{Karol Bartkiewicz\footnote{These authors have made equal contribution.}}
\affiliation{Institute of Physics, Adam Mickiewicz University,
61-614 Pozna\'n, Poland}

\author{Yu-xi Liu}
\affiliation{Frontier Research System, Institute of Physical and
Chemical Research (RIKEN), Wako-shi 351-0198, Japan}
\affiliation{CREST, Japan Science and Technology Agency (JST),
Kawaguchi, Saitama 332-0012, Japan}

\author{Adam Miranowicz}
\affiliation{Institute of Physics, Adam Mickiewicz University,
61-614 Pozna\'n, Poland} \affiliation{SORST Research Team for
Interacting Carrier Electronics, 4-1-8 Honmachi, Kawaguchi,
Saitama 331-0012, Japan} \affiliation{Graduate School of
Engineering Science, Osaka University, Toyonaka, Osaka 560-8531,
Japan}

\begin{abstract}
We investigate quantum teleportation through dissipative channels
and calculate teleportation fidelity as a function of damping
rates. It is found that the average fidelity of teleportation and
the range of states to be teleported depend on the type and rate
of the damping in the channel. Using the fully entangled fraction,
we derive two bounds on the damping rates of the channels: one is
to beat the classical limit and the second is to guarantee the
non-existence of any other copy with better fidelity. Effect of
the initially distributed maximally entangled state on the process
is presented; and the concurrence and the fully entangled fraction
of the shared states are discussed. We intend to show that prior
information on the dissipative channel and the range of qubit
states to be teleported is helpful for the evaluation of the
success of teleportation, where success is defined as surpassing
the fidelity limit imposed by the fidelity  of  1-to-2 optimal
cloning machine for the specific range of qubits.
\end{abstract}
\date{\today}
\pagestyle{plain} \pagenumbering{arabic} \maketitle

\section{Introduction}

The quantum state of a system can be transmitted from a location
to a distant one using only classical information provided that a
quantum channel exists between the sender and the receiver.
Sharing entangled states between the two parties opens the
necessary quantum channel \cite{HorodeckiReview}. Research in
quantum state transfer \cite{cloning}, especially the quantum
teleportation \cite{Bennett}, has emerged as one of the major
research areas of theoretical and experimental quantum mechanics.
Various discussions and criteria have appeared about the
evaluation of the state transfers under ideal and imperfect
conditions \cite{Holger}. In a perfect scheme, the shared
entangled state is a maximally entangled state (MES) enabling
perfect quantum state transfer. However, in practice, entanglement
is susceptible to local interactions with the environment, which
can result in loss of coherence. In this article, we study the
teleportation of qubits through damping channels.

We consider quantum state transfer as an operation, such as
cloning and teleportation, which beats the classical limits on
measurement and transmission. The resemblance of two quantum
states and the properties of quantum state transfer (teleportation
and cloning) are quantified by the fidelity $F(|\psi_{{\rm
in}}\rangle)=\langle\psi_{{\rm in}}|\hat{\rho}_{{\rm
out}}|\psi_{{\rm in}}\rangle$, which measures the overlap of the
states $|\psi_{\rm in}\rangle$ to be teleported (cloned) and the
output state with the density operator $\hat{\rho}_{{\rm out}}$.

A qubit state to be teleported $|\psi_{{\rm
in}}\rangle=\alpha|0\rangle+\beta|1\rangle$ with
$|\alpha|^{2}+|\beta|^{2}=1$ can be represented on a Bloch sphere
as
\begin{equation}\label{ktyu}
|\psi_{\rm in} \rangle=\cos(\delta/2)\E^{i\gamma}|0 \rangle
+\sin(\delta/2)|1\rangle,
\end{equation}
where $\delta$ and $\gamma$ are the polar and azimuthal angles,
respectively. Since this state is generally unknown, it is more
appropriate to calculate the average of the fidelity
$F(|\psi_{{\rm in}}\rangle)$ over all possible states $|\psi_{{\rm
in}}\rangle$ to quantify the process. This average fidelity
$F=\overline{\langle\psi_{{\rm in}}|\hat{\rho}_{{\rm
out}}|\psi_{{\rm in}}\rangle}$ \cite{Popescu} can be calculated as
\begin{eqnarray}\label{a1}
F=\frac{1}{4\pi}\int_{0}^{2\pi}d\gamma\int_{0}^{\pi}d\delta
F(\delta,\gamma)\sin\delta,
\end{eqnarray}
where the $4\pi$ is the solid angle.

The relation between the teleportation fidelity and the degree of
entanglement shared by the parties has been studied by many
researchers (e.g. in
\cite{Bennett,Holger,Horodecki,Popescu,Grosshans,Band,Carlo,Gordon}
and others cited in \cite{HorodeckiReview}) and it has been shown
that (i) less entangled quantum channel reduces the fidelity and
the range of states, which can be teleported \cite{Bennett}, (ii)
for the standard teleportation scheme, the maximum attainable
average fidelity is simply related to the fully entangled fraction
of a bipartite entangled state \cite{Horodecki}, and (iii) some
mixed states, which do not violate the Bell inequalities, can
still be used for teleportation \cite{Popescu}. On the other hand,
only a few studies are directed to the relation between the
fidelity of teleportation and the type and strength of the damping
in the quantum channel. That is the topic of the present study.

According to the definition of teleportation as stated by Bennett
{\em et al.} \cite{Bennett}, in the process of quantum
teleportation, one can construct an exact replica of the original
unknown quantum state with the cost of destroying the original
state. Therefore, to call a quantum state transfer operation as
quantum teleportation, the process should not only generate output
states with better qualities than what can be done classically but
also obey the no-cloning theorem \cite{Zurek}. Defining a
teleportation operator $\hat{U}_{\rm tel}$, which can be
implemented in a standard quantum circuit (see, e.g.,
\cite{Nielsen}) with an input state $\hat{\rho}_{\rm
in}=\hat{\rho}_{a}$ and a shared entangled state $\hat{\rho}_{\rm
ent}=\hat{\rho}_{b,c}$, the output state $\hat{\rho}_{\rm out}$ is
written as
\begin{equation}\label{cc1}
\hat{\rho}_{\rm out}={\rm Tr}_{\rm in,a}[\hat{U}_{\rm
tel}\hat{\rho}_{\rm in} \otimes\hat{\rho}_{\rm ent}
\hat{U}^{\dagger}_{\rm tel}].
\end{equation}
If the teleportation process is ideal then $\hat{\rho}_{\rm out}=
\hat{\rho}_{\rm in} $ implying a fidelity value of unity. However,
in practical applications, this is not the case due to the
presence of noise which may be due to (i) noisy sources of
$\hat{\rho}_{\rm in}$ and $\hat{\rho}_{\rm ent}$, (ii) noisy
entanglement distribution channel, (iii) noisy measurements and
unitary operators, and (iv) an eavesdropper who attempts to clone
$\hat{\rho}_{\rm in}$. Since, in general one cannot be sure of
which of the above are the reason, all the noise in the process
should be attributed to an eavesdropper in order to assess the
security whenever quantum teleportation is to be used as a means
of secure communication. This assessment to quantify the process
should be done according to the definition of the teleportation
given above. That is, one should check to see whether $F$ in Eq.
(\ref{a1}) satisfies the conditions of (i) beating the classical
limit, and (ii) obeying the no-cloning.

The linearity of quantum mechanics forbids the exact cloning of an
unknown quantum state, however, if one allows discrepancies
between the original quantum state and its copy, then it is
possible to devise a scheme that can produce clones and copies of
a given unknown state with the highest resemblance to the original
one \cite{Buzek,Gisin,Buzek1,Bruss1} (for reviews see
\cite{cloning}). This is known as the {\em optimal cloning}, where
with the increasing number of clones (copies), the resemblance to
the original state decreases. It has been shown that for a
state-independent universal cloning machine the relation between
the optimum fidelity $F$ of each copy and the number $M$ of copies
is given by $F=(2M+1)/(3M)$. In classical situations, one can make
infinite number of copies ($M\rightarrow\infty$) of a given state
resulting in a fidelity $F=2/3$, which is the best one can do with
classical operations. On the other hand, when $M=2$, the universal
cloning machine has an optimum fidelity of $F=5/6$
\cite{Buzek,Gisin,Buzek1,Bruss1}.

Combining the above information on teleportation and cloning, one
can infer that a teleportation process beats the classical limit
if $F>2/3$, and obeys the no-cloning requirement if $F>5/6$
\cite{Buzek,Gisin,Buzek1,Bruss1}. If this is assured, then there
is no any other copy of the output state with better fidelity,
therefore, the teleportation process is secure. It is noteworthy
that this is true if and only if the quantum state
$\hat{\rho}_{\rm in}$ is completely unknown to the eavesdropper.
In some cases, $\hat{\rho}_{\rm in}$ may be prepared in a state
that is selected from a known ensemble of states. If the
eavesdropper has this a priori knowledge about $\hat{\rho}_{\rm
in}$, a state dependent cloner which can perform better than the
optimal universal one can be constructed. Thus, the fidelity
constraint imposed on teleportation due to no-cloning condition
will become much stricter.

Quality of the shared entangled state is a good criterion to
quantify the reliability of the quantum teleportation. Bennett
{\em et al. } \cite{Bennett2} and, in general case, Horodecki {\em
et al. } \cite{Horodecki,Badziag} (for a review see
\cite{HorodeckiReview}) have shown that for a shared bipartite
entangled state $\hat{\rho}_{\rm ent}$ to be useful for quantum
teleportation, its fully entangled fraction $f_{{\rm ent}}$,
defined by \cite{Bennett1}
\begin{equation}\label{dd1}
f_{{\rm ent}}={\max_{\Phi}}~\langle\Phi|\hat{\rho}_{{\rm
ent}}|\Phi\rangle,
\end{equation}
must be greater than $1/2$. In Eq. (\ref{dd1}), maximum is taken
over all MES $|\Phi\rangle$. It has also been shown that, the
maximum achievable teleportation fidelity $F$ is related to
$f_{{\rm ent}}$ by \cite{Horodecki}
\begin{equation}\label{ee1}
 F=\frac{2f_{{\rm ent}}+1}{3}.
\end{equation}
States with $f_{{\rm ent}}\leq1/2$ cannot be used directly for
teleportation unless they are enhanced through filtering to
satisfy $f_{{\rm ent}}>1/2$. Choosing the boundary value of
$f_{{\rm ent}}=1/2$ gives a teleportation fidelity of $F=2/3$
which is the boundary between classical and quantum state
transfer. That is if $f_{{\rm ent}}\leq1/2$ and hence $F\leq2/3$,
then the same operation can be done classically. According to this
definition, if in a process $F>2/3$ is achieved then it can be
called quantum teleportation. On the other hand, the discussion on
cloning in the former paragraphs implies that one can make
infinite number of copies of a qubit with a fidelity of $2/3$
which violates the original definition of quantum teleportation
given by Bennett {\em et al. } \cite{Bennett}. Here again arises
the question of achievable teleportation fidelity, which
guarantees better than classical teleportation and surpasses the
no-cloning limit.

\begin{figure}[]
\epsfxsize=10cm \epsfbox{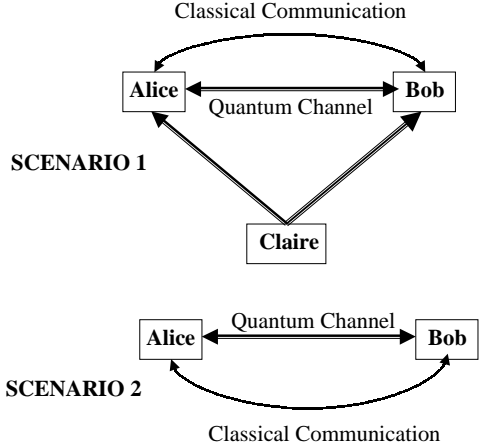} \caption[]{Teleportation
scenario 1, where both qubits of $\hat{\rho}_{\rm ent}$ are
affected by the channel, and scenario 2, where only one of the
qubits is affected. Quantum channel is formed by the shared
entangled state. } \label{fig1}
\end{figure}

The problem studied in this paper can be formulated as follows:
Alice and Bob are far from each other, and they share an entangled
quantum state $\hat{\rho}_{\rm en}$. The entangled state is
prepared either by a third party, say Claire, and delivered to
Alice and Bob (scenario 1: two-qubit affected scenario) or
prepared by Alice and one of the qubits is sent to Bob and the
other is kept with her (scenario 2: one-qubit affected scenario)
as shown in Fig.~\ref{fig1}. The only manipulations that Alice and
Bob are allowed to do is local quantum operations and classical
communications. Now suppose that Alice wants to transfer the
quantum state represented by the qubit state $|\psi_{\rm
in}\rangle$ to Bob and the entangled state is distributed through
a dissipating channel. Then, how does the dissipation of the
channel affect the entanglement properties of the distributed
entangled state, and hence what is its effect on the transferred
quantum state? What is the allowable amount of dissipation that
does not affect the security of quantum state transfer?

In this paper, we derive the damping rates of quantum channels at
which a quantum state transfer that overcomes the classical
counterpart can be realized. In the same way, conditions, which
guarantee a secure quantum teleportation, are also derived. We
study the effect of noise on the range of qubits that can be
teleported accurately. The noisy channels, including amplitude
damping channel, phase damping channel and depolarizing channel,
and the effects of these noisy channels on the distributed
entanglement and teleportation process are studied in Sec. II and
III. And finally, Sec. IV includes a brief summary and conclusion
of this study.

\section{Effect of Damping Channels on Entanglement and Teleportation}

We consider the two scenarios shown in Fig.~\ref{fig1}. In the
first scenario, the qubits of the initial MES are distributed
through two channels, which may or may not have the same damping
properties. On the other hand, in the second scenario, only one of
the qubits of the initial MES is distributed through the damping
channel. In the following, we give analytical expressions, which
show how a given state is affected when transmitted through noisy
channels causing amplitude damping, phase damping or
depolarization. Initial MES that are considered in this study are
the Bell states
\begin{eqnarray}
 |\psi^{\pm}\rangle&=&\frac{1}{\sqrt{2}}(|01\rangle\pm|10\rangle),
 \nonumber \\
 |\phi^{\pm}\rangle&=&\frac{1}{\sqrt{2}}(|00\rangle\pm|11\rangle).
\label{N1}
\end{eqnarray}
We derive the bounds for the damping rate of the channel to
satisfy the quantum teleportation conditions discussed in the
previous section. In the following, we assume that there is no a
priori information on $\hat{\rho}_{\rm in}$, therefore the optimal
universal cloning machine which imposes $F>5/6$ is considered.

\subsection{Amplitude Damping Channel}

The evolution of environment (denoted by subscript $e$) and a
system (subscript $a$ or, equivalently, $b$) with the states
$|0\rangle$ and $|1\rangle$ is defined by the following
transformation in the presence of the amplitude damping channel
(ADC) \cite{Preskill}:
\begin{eqnarray}\label{N01}
|0\rangle_{a}|0\rangle_{e}&&\rightarrow|0\rangle_{a}|0\rangle_{e}\nonumber\\
|1\rangle_{a}|0\rangle_{e}&&\rightarrow\sqrt{q}
|1\rangle_{a}|0\rangle_{e}+\sqrt{p}|0\rangle_{a}|1\rangle_{e}
\end{eqnarray}
where $q\equiv1-p$. This transformation implies that a system with
an excited state makes a transition to the ground state with a
probability $p$ and emits a photon to the environment which makes
a transition to the excited state. When the system is initially in
the ground state, there is no transition.

\subsubsection{Input Bell states $|\psi^{\pm}\rangle$}

If both of the qubits in the Bell states $|\psi^{\pm}\rangle$ are
transmitted through an ADC (scenario 1), then using Eq.
(\ref{N01}), we can write the state at the output of the channel
as
\begin{eqnarray}\label{N02}
|\Psi^{\pm}\rangle_{abe_{1}e_{2}}
&=&\frac{1}{\sqrt{2}}[(\sqrt{q_{b}}|01\rangle
_{ab}\pm\sqrt{q_{a}}|10\rangle
_{ab})|00\rangle _{e_{1}e_{2}}~~~\nonumber\\
&&+(\sqrt{p_{b}}|01\rangle _{e_{1}e_{2}}\pm\sqrt{p_{a}}|10\rangle
_{e_{1}e_{2}})|00\rangle _{ab}],
\end{eqnarray}
where we assumed that channels have different damping rates
denoted by $p_{a}$, $p_{b}$ and, for simplicity, we denote
$q_a\equiv 1-p_a$, $q_b\equiv 1-p_b$. If we assume $p_{a}=p_{b}=p$
and the environment is not monitored (unwatched channel), the
shared state between Alice and Bob at the outputs of the channels
can be found by tracing out the environment variables resulting in
$\hat{\rho}^{\pm}_{ab}=q|\psi^{\pm}\rangle_{ab\;ab}\langle\psi^{\pm}|
+p|00\rangle_{ab\;ab}\langle00|$. It is seen that the MES survives
with a probability of $q$. On the other hand, if the environment
is monitored, Alice and Bob proceed with the protocol if no photon
is detected in the environment implying they have a MES, and they
do nothing when photon is detected in the environment.

If only one of the qubits (say, that for Bob) is sent through the
channel (scenario 2), the damping in the channel affects only that
part. If the channel is watched and no photon is detected in the
environment, the state that is shared between Alice and Bob
becomes
\begin{eqnarray}\label{N03}
|\Psi^{\pm}\rangle_{ab}=\frac{1}{\sqrt{2-p_{b}}}(\pm\sqrt{q_{b}}
|01\rangle _{ab}+|10\rangle _{ab}).
\end{eqnarray}
For an unwatched channel, the shared state is given as
\begin{equation}\label{N04}
\hat{\rho}^{\pm}_{ab}=\frac{1}{2}[(2-p_{b})|\Psi^{\pm}\rangle_{ab\,ab}
\langle\Psi^{\pm}| +p_{b}|00\rangle_{ab\,ab}\langle00|].
\end{equation}
It is clearly seen that if only one qubit of the initial MES is
sent through the ADC, the shared state between the parties is no
longer a MES.

Using Eq. (\ref{N02}), one can find that for scenario 1, the fully
entangled fraction is \cite{Band}
\begin{eqnarray}\label{gg1}
f^{\pm}_{\rm ent,1}=\frac{1}{4}(\sqrt{q_{a}}+\sqrt{q_{b}})^{2}
\end{eqnarray}
if $p_{a}\ge \frac12 (q_{b}+\sqrt{1+2p_{b}-3p_{b}^2})$ is
satisfied, otherwise it becomes
\begin{eqnarray}\label{gg1a}
f^{\pm}_{\rm ent,2}&=&\frac{1}{4}(p_{a}+p_{b}).
\end{eqnarray}
If we assume that both channels have the same damping properties
that is $p_{a}=p_{b}=p$, the fully entangled fraction is found as
\begin{equation}\label{gg1abc}
 f^\pm_{{\rm ent}}=
\left\{
\begin{array}{cc}
 q ~~~~~~~~~~~~~&\text{if $p\le 2/3$;} \\
   &          \\
 p/2 ~~~~~~~~~~~~~& \text{if $p> 2/3$. }
\end{array}\right.
\end{equation}
For scenario 2, where $p_{a}=0$, $f^{\pm}_{\rm ent}$ becomes
$f^{\pm}_{\rm ent}=\frac{1}{4}(1+\sqrt{q_{b}})^{2}$ for all
$p_{b}$.

Imposing the condition $f_{{\rm ent}}>1/2$, which assures that a
quantum state operation beats the classical limit, gives the
relation $\sqrt{q_{a}}+\sqrt{q_{b}}>\sqrt{2}$ for scenario 1
\cite{Band}. Taking $p_{b}$ as a variable, it can be found that
$p_{a}<2(\sqrt{2}-1)$ and $p_{b}\leq p_{a}-2+2\sqrt{2q_{a}}$ must
be satisfied simultaneously \cite{Band}. Similarly for the case
$p_{a}=p_{b}=p$, one can find that classical limit can be beaten
only when $p<1/2$. For scenario 2, it can easily be shown that
$p_{b}<2(\sqrt{2}-1)$ must be satisfied. If the channels are
watched but no photon is detected, then $f_{{\rm ent}}>1/2$ can
always be satisfied provided that $p_{a}<1\wedge p_{b}<1$ and
$p_{b}<1$ for scenarios 1 and 2, respectively.

If the conditions given in the above paragraph for unwatched
channels are satisfied, one can only be sure that the operation is
a quantum one with fidelity $F>2/3$, however, cannot be sure about
the security of the process, which requires $F>5/6$ according to
$1\rightarrow2$ cloning condition. Then solving Eq. (\ref{ee1})
for $f_{{\rm ent}}$ to satisfy $F>5/6$, we find that
\begin{eqnarray}\label{gg1b}
f_{{\rm ent}}>\frac{3}{4}
\end{eqnarray}
must be satisfied for the fully entangled fraction. Imposing this
condition on the shared entangled state between the two parties
results in a much tighter condition on the channel damping rates,
which can be summarized as follows
\begin{eqnarray}
&& f^{\pm}_{\rm ent}>\frac{3}{4}~ \text{if}\nonumber\\
\nonumber\\
&&\left\{
\begin{array}{cc}
p<\frac{1}{4}&\text{scenario 1} \\
   &        \text{for } p_{b}=p_{a}=p,\\
   & \\
p_{b}<p_{a}-3+2\sqrt{3q_{a}}&\text{scenario 1}\\
p_{a}\leq 2\sqrt{3}-3\text{ or vice versa}&\text{for $p_{b}\neq p_{a}$},\\
&  \\
p_{b}< 2\sqrt{3}-3& \text{scenario 2}.
\end{array}\right.
 \end{eqnarray}

\subsubsection{Input Bell states $|\phi^{\pm}\rangle$}

In the scenario 1, when both qubits of $|\phi^{\pm}\rangle$ are
sent through the damping channels, the shared state between Alice
and Bob for watched and unwatched channel are found as
\begin{eqnarray}\label{N05}
|\Phi^{\pm}\rangle_{ab}=\frac{|00\rangle
_{ab}\pm\sqrt{q_{a}q_{b}}|11\rangle
_{ab}}{\sqrt{1+q_{a}q_{b}}},
\end{eqnarray}
and
\begin{eqnarray}\label{N06}
\hat{\rho}^{\pm}_{ab}=&&\frac{1}{2}
\left[(1+q_{a}q_{b})|\Phi^{\pm}\rangle_{ab\;ab}
\langle\Phi^{\pm}|\right. \nonumber\\
&&\left. +q_{b}p_{a}|01\rangle_{ab\;ab}\langle01|
+p_{b}q_{a}|10\rangle_{ab\;ab}\langle10|\right. \nonumber\\
&&\left. +p_{a}p_{b}|00\rangle_{ab\;ab}\langle00| \right],
\end{eqnarray}
respectively, where we have considered that no photon is detected
in the environment for the watched channel case. From these
equations, it is seen that a MES survives with a non-zero
probability iff $p_{a}=p_{b}=0$.

When only one of the qubits (say again, Bob's qubit) of the MES is
propagated through the ADC, the shared state between Alice and Bob
is not maximally entangled unless $p_{b}=0$ for both watched and
unwatched channels as can be seen in the following expressions
given, respectively, for watched and unwatched channels
\begin{eqnarray}\label{N07}
|\Phi^\pm\rangle_{ab}=\frac{1}{\sqrt{2-p_{b}}}(|00\rangle
_{ab}\pm\sqrt{q_{b}}|11\rangle _{ab}),
\end{eqnarray}
and
\begin{equation}\label{N08}
\hat{\rho}^{\pm}_{ab}=\frac{1}{2}\left[(2-p_{b})|\Phi^\pm\rangle_{ab\;ab}
\langle\Phi^\pm|+ p_{b}|10\rangle_{ab\;ab}\langle 10|\right].
\end{equation}
Then the fully entangled fraction of the shared state, when the
channel is not watched, is found as
\begin{equation}\label{gg1abd}
 f^{\pm}_{\rm ent}=\frac{1}{4}[p_{a}p_{b}+(1+\sqrt{q_{a}q_{b}})^{2}],
\end{equation}
which reduces to
\begin{equation}\label{gg1abe}
 f^{\pm}_{\rm ent}=\frac{1}{4}
\left\{
\begin{array}{cc}
  2(p^{2}-2p+2)& \text{scenario 1,}\\
&        \text{for } p_{a}=p_{b}\equiv p,\\
   & \\
 (1+\sqrt{q_{b}}~)^{2}&\text{scenario 2}. \\
\end{array}\right.
 \end{equation}
It can easily be found from Eq. (\ref{gg1abe}) that, the condition
$f_{{\rm ent}}>1/2$ is satisfied for any $p$ in the range $p<1$
when both channels have the same damping rates in scenario 1; and
for $p_{b}<2(\sqrt{2}-1)$ in scenario 2. When the channels have
different damping rates, we can write using Eq. (\ref{gg1abd})
that $p_{a}p_{b}+(1+\sqrt{q_{a}q_{b}})^{2}>2$ must be satisfied to
beat the classical limit. Analytical solution for this is very
lengthy to give here. Instead, to give an idea on the relation
between $p_{b}$ and $p_{a}$ to satisfy the condition $f_{{\rm
ent}}>1/2$, we give some numerical values: when $p_{b}=1/2$,
$p_{a}<7/8$ and when $p_{b}=1/4$, $p_{a}$ must satisfy
$p_{a}<(6\sqrt{6}-13)/2$ to beat the classical limit. As it has
been pointed out by Bandyopadhyay \cite{Band}, scenario 1 can be
made to have higher $f_{{\rm ent}}$ than scenario 2 such that
$f_{{\rm ent}}>1/2$ is satisfied. This, in turn, implies that for
the state $|\phi^{\pm}\rangle$, one can let one of the qubits to
undergo a controlled dissipation if the information on the
dissipation of the other qubit in the other channel is available.

Looking at the condition $f_{{\rm ent}}>3/4$ for quantum
teleportation to surpass the no-cloning limit, we find the
following constraints on the damping rates of the ADC:
\begin{eqnarray}
&& f^\pm_{{\rm ent}}>\frac{3}{4}~ \text{if}\nonumber\\
\nonumber\\
&&\left\{
\begin{array}{cc}
 p<1-\frac{\sqrt{2}}{2}&\text{scenario 1} \\
   &         \text{for~} p_{a}=p_{b}\equiv p,\\
   & \\
p_{a}\leq 2\sqrt{3}-3&\text{scenario 1}\\
p_{b}\leq g(p_{a}) \text{ or vice versa}&\text{for $p_{a}\neq p_{b}$}{,}\\
&  \\
p_{a}< 2\sqrt{3}-3& \text{scenario 2,}
\end{array}\right.
 \end{eqnarray}
where $g(x)=(1-2x)^{-2}[-3+x(3+2x)+2\sqrt{(1-x)(2x^2-6x+3)}]. $
Contrary to the above case, a controlled dissipation cannot
increase $f_{{\rm ent}}$ above $3/4$.

\subsection{Phase Damping Channel}

A phase damping channel (PDC) affects an input state with the
following transformations \cite{Preskill}
\begin{eqnarray}\label{N05a}
|0\rangle_{a}|0\rangle_{e}&&\rightarrow\sqrt{q}|0\rangle_{a}|0\rangle_{e}
+\sqrt{p}|0\rangle_{a}|1\rangle_{e},\nonumber\\
|1\rangle_{a}|0\rangle_{e}&&\rightarrow\sqrt{q}|1\rangle_{a}|0\rangle_{e}
+\sqrt{p}|1\rangle_{a}|2\rangle_{e}.
\end{eqnarray}
In this channel, the energy of the information carrier is
conserved (no losses to environment), however the state of the
carrier is decohered.

\subsubsection{Input Bell states $|\psi^{\pm}\rangle$}

Bell states $|\psi^{\pm}\rangle$ evolve into
\begin{eqnarray}\label{N09}
\hat{\rho}^{\pm}_{ab}&=&\frac{1}{2}(1-q_{a}q_{b})
(|01\rangle_{ab\;ab}\langle01|+|10\rangle_{ab\;ab}\langle10|)\nonumber\\
&&~~~~~~~~~~~+q_{a}q_{b}|\psi^{\pm}\rangle_{ab\;ab}\langle\psi^{\pm}|,
\end{eqnarray}
when both qubits are sent through the unwatched channel. For the
limiting case, $p_{a}=1\vee p_{b}=1$, off-diagonal components of
the density matrix vanish resulting in a mixed state.  For a
watched channel with no photon detected in the environment, there
is a probability of $q_{a}q_{b}$ that the state observed is
$|\psi^{\pm}\rangle$.

On the other hand, when only one qubit is sent (scenario 2), the
probability that the MES survives becomes $q_{b}$ when the channel
is watched. When the channel is not watched, then the output
state, which is mixed and not a MES, can be found from Eq.
(\ref{N09}) by substituting $p_{a}=0$.

When the $f^{\pm}_{\rm ent}$ of the output state at the end of the
unwatched channels are calculated it is seen that
\begin{equation}\label{hh1abe}
 f^{\pm}_{\rm ent}=\frac{1}{2}
\left\{
\begin{array}{cc}
  1+q_{a}q_{b}& \text{scenario 1}\\
&          \text{for~} p_{a}\neq p_{b},\\
   & \\
p^{2}-2p+2& \text{scenario 1,}\\
&          \text{for~} p_{a}=p_{b}\equiv p,\\
   & \\
 2-p_{b}&\text{scenario 2}. \\
\end{array}\right.
 \end{equation}
Then we find that $f^{\pm}_{\rm ent}$ is always greater than $1/2$
provided that $p_{b}\neq 1 \wedge p_{a}\neq 1$ and $p\neq 1$ are
satisfied for both scenarios. Moreover, we find scenario 1 cannot
be made to have $f^{\pm}_{\rm ent}$ larger than scenario 2.

The no-cloning limit imposes the following conditions on the
allowable PDC rate:
\begin{eqnarray}
&& f^\pm_{{\rm ent}}>\frac{3}{4}~ \text{if}\nonumber\\
\nonumber\\
&&\left\{
\begin{array}{cc}
p<1-\frac{\sqrt{2}}{2}&\text{scenario 1} \\
   &          \text{for~} p_{a}=p_{b}\equiv p,\\
   & \\
 p_{b}<(1-2p_{a})/(2q_{a})&\text{scenario 1}\\
 p_{a}<1/2 \text{ or vice versa}&\text{for $p_{a}\neq p_{b}$}{,}\\
&  \\
 p_{b}< 1/2& \text{{scenario} 2{. }}
\end{array}\right.
 \end{eqnarray}

\subsubsection{Input Bell states $|\phi^{\pm}\rangle$}

When the input is $|\phi^{\pm}\rangle$, then the state at the
output of the channels becomes
\begin{eqnarray}\label{N10}
\hat{\rho}^{\pm}_{ab}&&=\frac{1}{2}(1-q_{a}q_{b})
(|00\rangle_{ab\;ab}\langle00|+|11\rangle_{ab\;ab}\langle11|)\nonumber\\
&&~~~~~~~~~~~+q_{a}q_{b}|\phi^{\pm}\rangle_{ab\;ab}\langle\phi^{\pm}|
\end{eqnarray}
for an unwatched channel for scenario 1. A comparison of this
output state with Eq. (\ref{N09}) reveals that the same
discussions and the conditions on the channel damping properties
are valid here, too.

\subsection{Depolarizing Channel}

When a qubit is sent through a depolarizing channel (DC), with a
probability $q\equiv 1-p$ it is intact, while with probability
$p~$ an error (bit flip error, phase flip error or both) occurs.
The transformation that characterizes this channel is
\cite{Preskill}
\begin{eqnarray}\label{N10a}
|0\rangle_{a}|0\rangle_{e}&&\rightarrow\sqrt{1-\frac{3p}{4}}~
|0\rangle_{a}|0\rangle_{e} \nonumber\\
&& +\sqrt{\frac{p}{4}}~\big(|1\rangle_{a}|1\rangle_{e}
+i|1\rangle_{a}|2\rangle_{e}+|0\rangle_{a}|3\rangle_{e}\big),
\nonumber\\
|1\rangle_{a}|0\rangle_{e}&&\rightarrow\sqrt{1-\frac{3p}{4}}|1\rangle_{a}
|0\rangle_{e} \\ &&
+\sqrt{\frac{p}{4}}~\big(|0\rangle_{a}|1\rangle_{e}
-i|0\rangle_{a}|2\rangle_{e}-|1\rangle_{a}|3\rangle_{e}\big).
\nonumber
\end{eqnarray}
In the DC, any given state $|\varphi\rangle$ evolves to an
ensemble of the four states $|\varphi\rangle$,
$\hat\sigma_{x}|\varphi\rangle$, $\hat \sigma_{y}|\varphi\rangle$
and $\hat\sigma_{z}|\varphi\rangle$ where $\sigma_{k}$ is the
Pauli operator. $p=1$ corresponds to complete depolarization where
each of the four states occur with equal probabilities.

If the input state to the channel is $|\psi^{\pm}\rangle$ or
$|\phi^{\pm}\rangle$ both qubits are sent through the channel then
with a probability of $(4-3p_{a})(4-3p_{b})/16$, this state is
conserved at the output of the channels if the channel is watched
and no photon is detected. For an unwatched channel with an input
$|\eta^{\pm}_{1,2}\rangle =
{|\psi^{\pm}\rangle,|\phi^{\pm}\rangle}$ for indices 1 and 2,
respectively, the output state can be written as
\begin{eqnarray}\label{N11} \hat{\rho}^{\pm}_{ab}&=&
\frac{1-q_{a}q_{b}}{4}(|\eta_{1,2}^{\mp}\rangle
\langle\eta_{1,2}^{\mp}|+|\eta_{2,1}^{-}\rangle\langle\eta_{2,1}^{-}|+
|\eta_{2,1}^{+}\rangle\langle\eta_{2,1}^{+}|)
\nonumber\\
&& + \frac{1+3q_{a}q_{b}}{4} |\eta_{1,2}^{\pm}\rangle\langle\eta_{1,2}^{\pm}|,
\end{eqnarray}
which becomes a mixture of Bell states with equal probability
$1/4$ when $p=1$. The effect of this channel on the input state
when only one of the qubits is sent through can be found simply by
substituting $p_{a}=0$. Imposing the criteria $f_{{\rm ent}}>1/2$
and $f_{{\rm ent}}>3/4$ on the state at the output of the channel
for both scenarios, we find the following ranges for damping rate
of the DC
\begin{eqnarray}\label{yyzccc}
&& f^{\pm}_{\rm ent}>\frac{1}{2}~ \text{if}\nonumber\\
\nonumber\\
&&\left\{
\begin{array}{cc}
p<1-\sqrt{3}/3&\text{scenario 1} \\
   &        \text{for}~ p_{b}=p_{a}=p,\\
   & \\
p_{a}<\frac{2-3p_{b}}{3q_{b}};~~ p_{b}<2/3 &\text{scenario 1}\\
 &\text{for~ $p_{b}\neq p_{a}$},\\
&  \\
p_{b}< 2/3& \text{scenario 2}
\end{array}\right.
 \end{eqnarray}
and
\begin{eqnarray}\label{xc12}
&& f^{\pm}_{\rm ent}>\frac{3}{4}~ \text{if}\nonumber\\
\nonumber\\
&&\left\{
\begin{array}{cc}
p<1-\sqrt{6}/3&\text{scenario 1} \\
   &       \text{for}~ p_{a}=p_{b}\equiv p,\\
   & \\
p_{a}<\frac{1-3p_{b}}{3q_{b}};~~ p_{b}<1/3 &\text{scenario 1}\\
 &\text{for~ $p_{b}\neq p_{a}$},\\
&  \\
p_{b}< 1/3& \text{scenario 2}.
\end{array}\right.
 \end{eqnarray}

\subsection{Concurrence and fully entangled fraction}

Fully entangled fraction $f_{{\rm ent}}$, given by (\ref{dd1}),
can be regarded as a measure of entanglement when the quantum
channel is in a pure state and it is related to the Wootters
concurrence $C$ \cite{Wootters} through the relation $f_{{\rm
ent}}=(1+C)/2$. However, when the quantum channel is in a mixed
state, $f_{{\rm ent}}$ can no longer be used as a measure of
entanglement. This is due to the fact that entanglement cannot be
increased by local quantum operations and classical
communications, but the fully entangled fraction $f_{{\rm ent}}$
can be increased as shown by Bandyopadhyay \cite{Band} and
Badzi\c{a}g {\em et al. } \cite{Badziag}.

In the following, we present the dependence of concurrence on the
properties of the unwatched damping channels in both scenarios
introduced previously and discuss the relation between $f_{{\rm
ent}}$ and concurrence so that we can assure a quantum state
transfer and a secure quantum teleportation. As defined by
Wootters \cite{Wootters}, concurrence of a mixed state $\varrho$
is given by $C=\max(0,\lambda_{1}-\lambda_{2}
-\lambda_{3}-\lambda_{4})$, where \{$\lambda_{i}$\} are the
square-roots of the eigenvalues, in decreasing order, of the
non-Hermitian matrix $\varrho\tilde{\varrho}$ with
$\tilde{\varrho}= \hat \sigma_{y}\otimes\hat
\sigma_{y}\varrho^{*}\hat \sigma_{y} \otimes\hat \sigma_{y}$,
where Pauli matrices act on Alice and Bob qubit respectively and
(*) stands for complex conjugation.

For the ADC, we observe that, in the first scenario, the relations
between the concurrence and the damping parameter are different
for the initial entangled states $|\psi^{\pm}\rangle$ and
$|\phi^{\pm}\rangle$. The concurrence of the shared state at the
output of the channels for the input state $|\psi^{\pm}\rangle$ is
given as $C'=\sqrt{q_{a}q_{b}}$. And the concurrence for
$|\phi^{\pm}\rangle$ becomes $C''=(1-\sqrt{p_{a}p_{b}})C'$. When
both channels have the same damping rate $p_{a}=p_{b}\equiv p$. It
is seen that while $C'$ decreases linearly with $p$, $C''$
decreases with $p^{2}$. On the other hand, for scenario 2, both
initial states show the same tendency, which is given as
$C'=C''=\sqrt{q}$.

\begin{figure}[]
\hspace*{32mm}(i) \hspace*{32mm}(ii)
 \epsfxsize=3.9cm\epsfbox{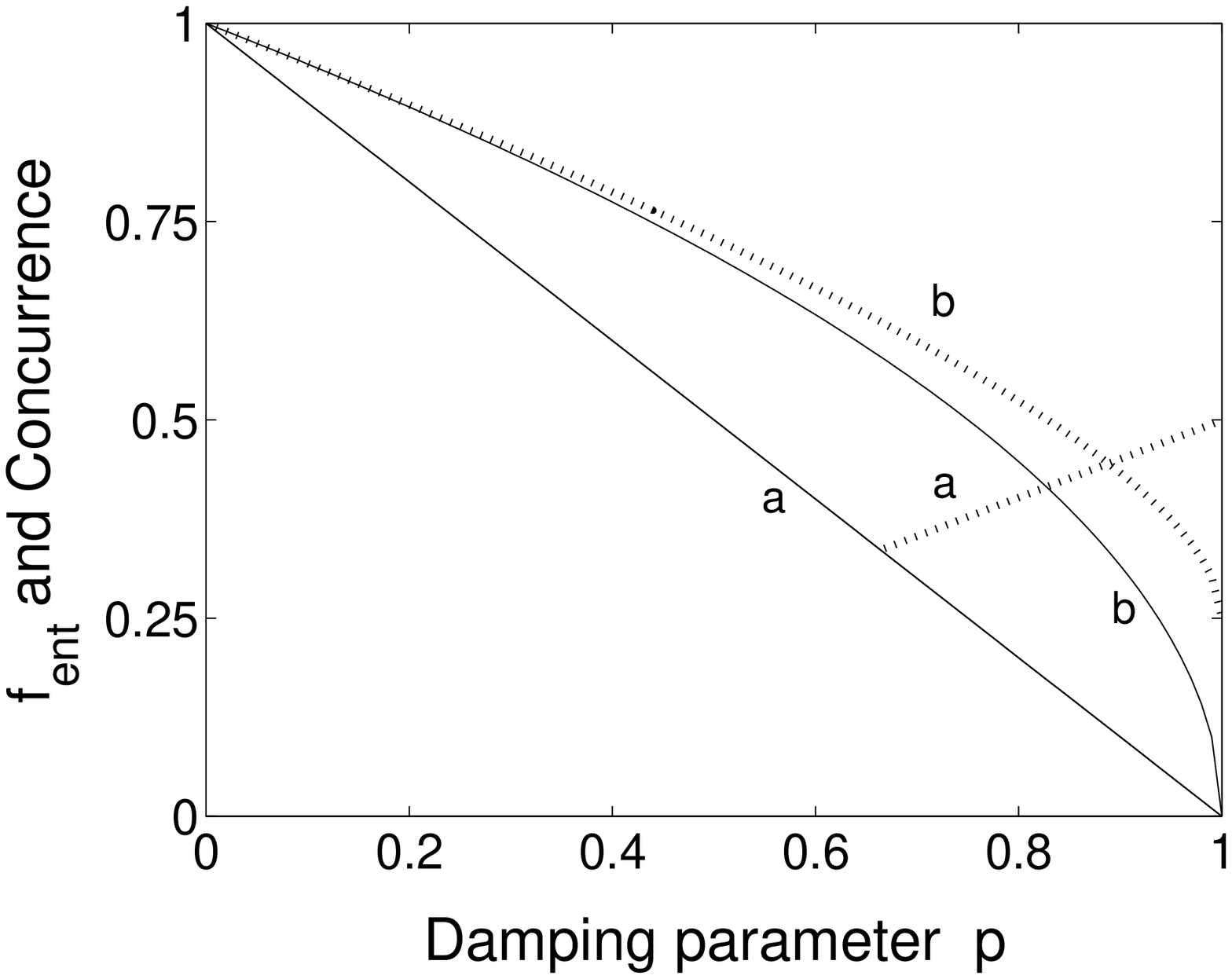}\hspace*{0mm}
 \epsfxsize=3.6cm\epsfbox{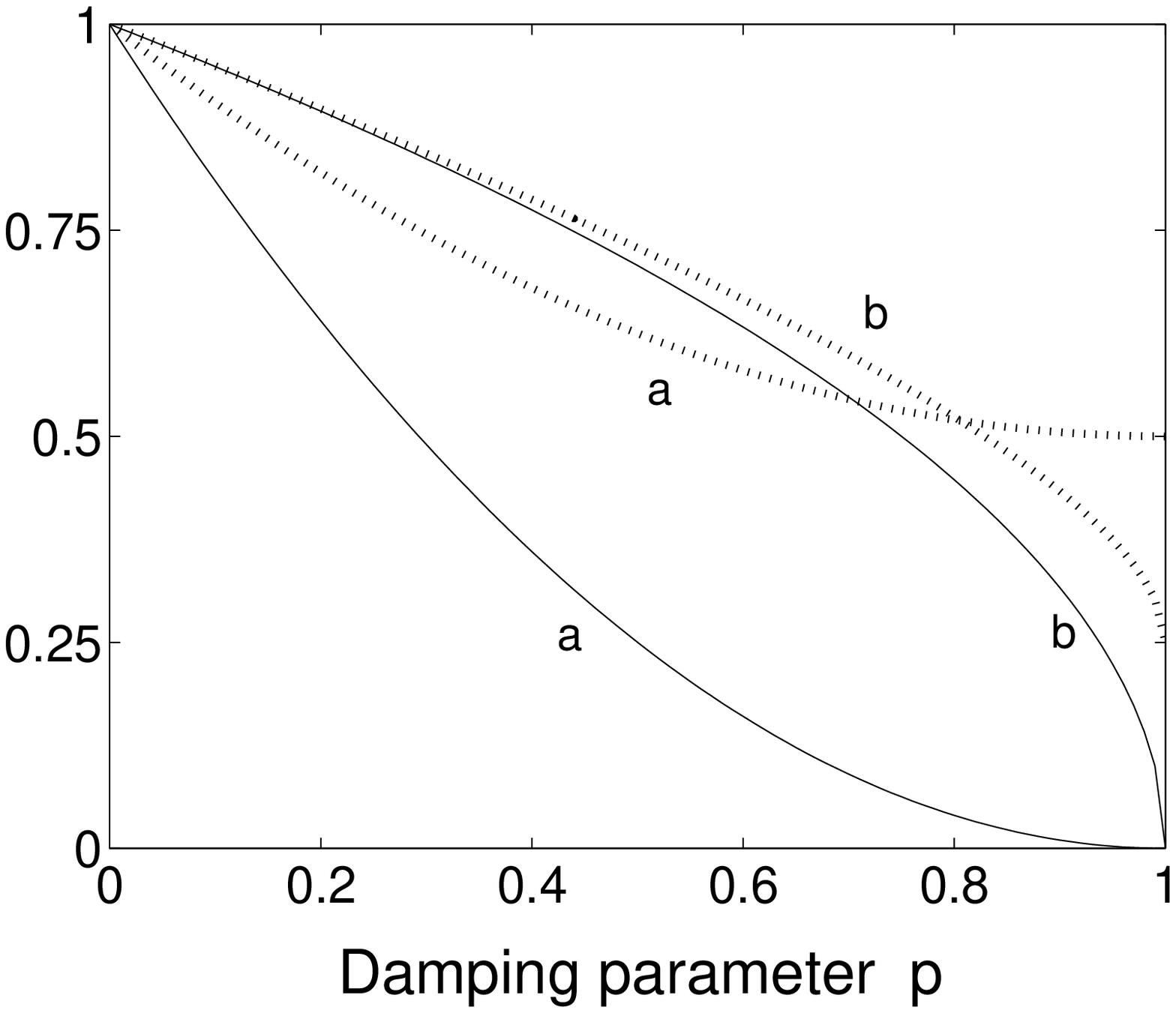}\vspace{0mm}\\
\hspace*{30mm}(iii) \hspace*{32mm}(iv)
 \epsfxsize=3.9cm\epsfbox{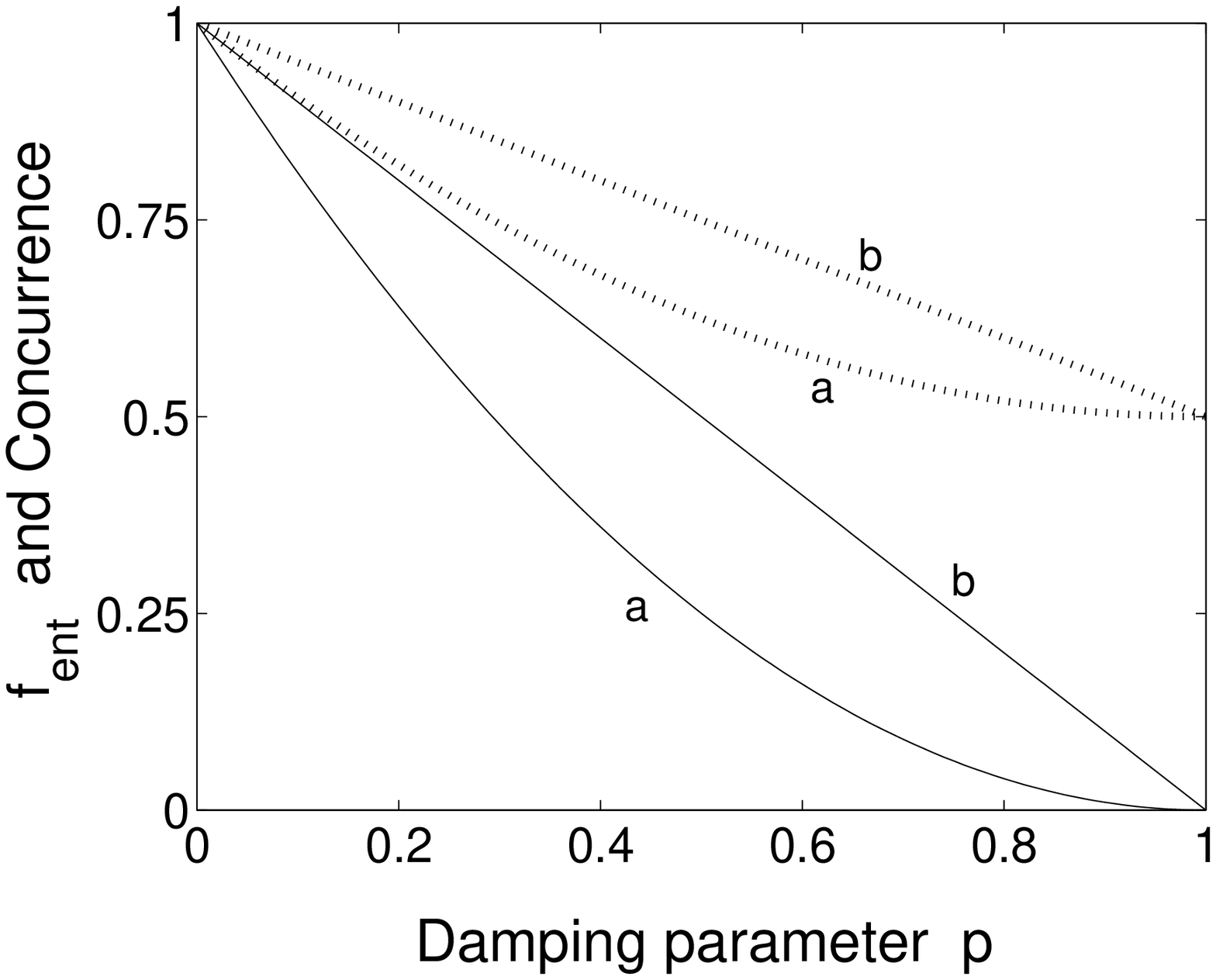}\hspace*{0mm}
 \epsfxsize=3.6cm\epsfbox{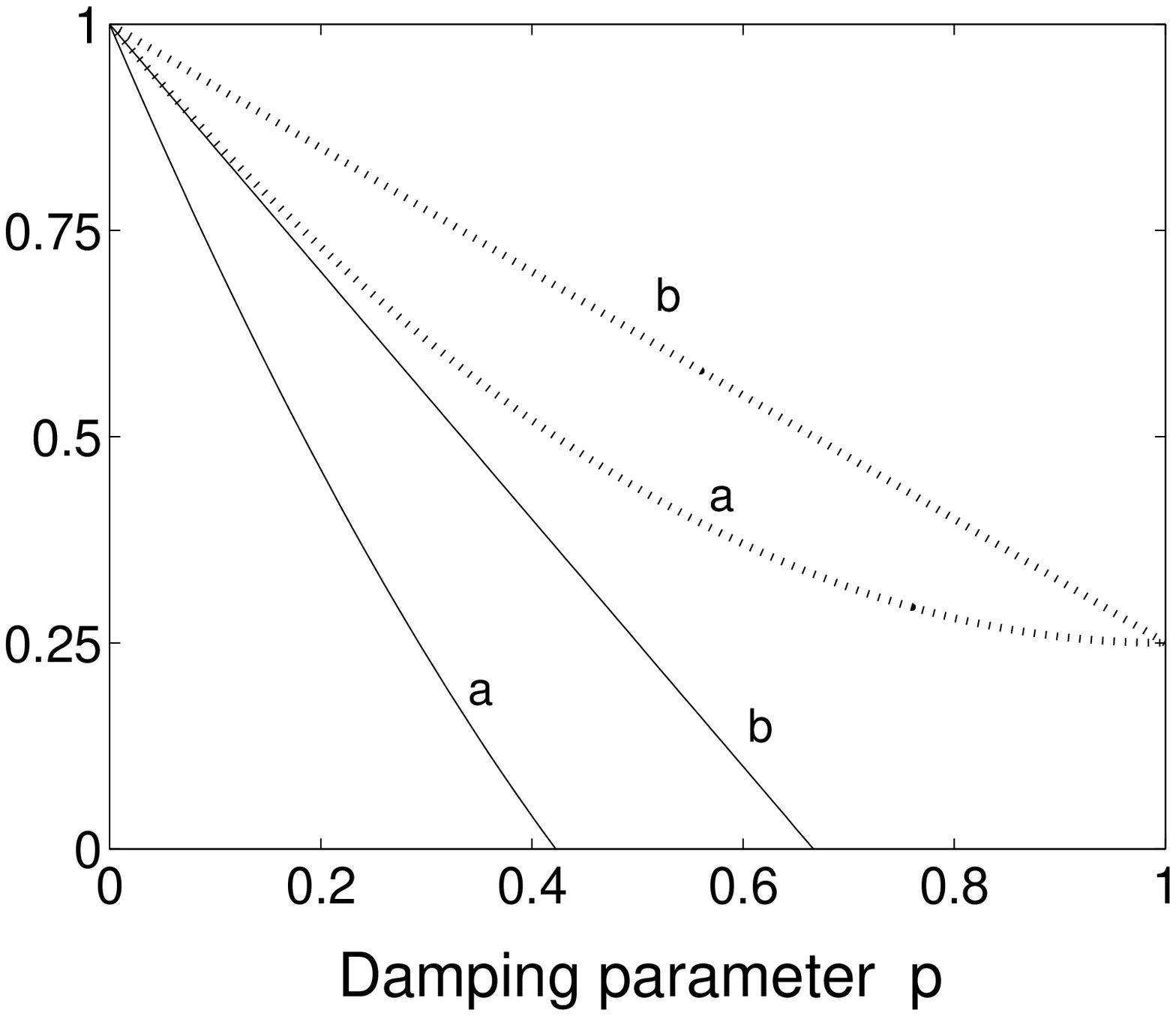}\vspace{0mm}
\caption[]{Comparison of concurrence (solid curves) and fully
entangled fraction (dashed curves) for the ADC, when the initial
entangled state is $|\psi^\pm\rangle$ (i) and $|\phi^\pm\rangle$
(ii), as well as the PDC (iii) and the DC (iv) for scenarios 1
(curves a) and 2 (curves b). Note that for the PDC and DC results
are independent of initial entangled state. } \label{fig2}
\end{figure}

In the cases of the PDC and DC, $C \equiv C'=C''$. For the PDC,
concurrence is found as $C=q_{a}q_{b}$ for the first scenario. The
expression for the second scenario can be found by taking
$p_{a}=0$ and $p_{b}=p$. Although the expressions found for
concurrence for the ADC and PDC are valid for all values of
$p_{a}$ and $p_{b}$ in the range of $[0,1]$, the expressions for
concurrence in the case of the DC are valid only for a limited
range of damping rates. For example, for the second scenario,
concurrence is found as $C=1-3p/2$ provided that $p<2/3$,
otherwise it is zero implying a separable state. For the first
scenario when both DCs have the same damping rate, concurrence is
given as $C=1+3p(p-2)/2$ when $p\leq 1-\sqrt{3}/3$, otherwise
$C=0$. When the damping rates are different, we find that
$C=(3q_{a}q_{b}-1)/2$ provided that $p_{b}< 2/3$ and $p_{a}<1-1/(3
q_{b})$ are satisfied simultaneously, otherwise $C=0$. It is seen
from Fig.~\ref{fig2} that $f_{{\rm ent}}$ is always $\leq 1/2$ for
$C=0$. And even a very small amount of entanglement shifts the
process from classical to quantum regime.

\section{range of qubits \protect{for accurate teleportation}}

In this section, we analyze the effect of the noise in the system
on the range of qubits that can be teleported with a desired
fidelity value. In order to show how the a priori information on
the ensemble from which $\hat{\rho}_{{\rm in}}$ is prepared
affects the fidelity criterion on secure teleportation, we will
consider an optimal one-to-two phase-covariant cloning machine
(PCCM) \cite{Bruss2,Fiurasek} in comparison with the universal
cloning machine. We will assume that the states to be teleported
are chosen from the whole set of qubit states with a fixed and
specified polar angle $\delta$ in the Bloch sphere. An
eavesdropper, who knows $\delta$, can use the optimal (one-to-two)
PCCM for which the cloning fidelity is given by
\cite{Fiurasek,Du}:
\begin{eqnarray}
  F'(\delta) = \frac12 \sin^2\frac{\delta+\kappa\pi}{2}
  +\cos^4\frac{\delta-\kappa\pi}{2}
  +\frac{\sqrt{2}}4 \sin^2\delta \hspace{8mm}
 \\
  = \frac18\left[5+\sqrt{2}+2\cos(\delta+\kappa)-(\sqrt{2}-1)\cos(2\delta)\right],
    \nonumber
\label{new1}
\end{eqnarray}
where $\kappa=[[2\delta/\pi]]$, i.e., $\kappa=0$ for
$0\le\delta<\frac{\pi}2$ and $\kappa=1$ for
$\frac{\pi}2\le\delta\le\pi$. Note that fidelity $F'(\delta)$ for
any $\delta$ is greater than the fidelity of the optimal universal
cloning machine \cite{Buzek}, given by $F=5/6$. For the qubit
states on the equator of the Bloch sphere ($\delta=\pi/2$), the
optimal PCCM prepares clones with fidelity
$F'(\pi/2)=\frac14(2+\sqrt{2})$. On the other hand, when the
states are close to the poles that is in the neighborhood of
$|1\rangle$ or $|0\rangle$  in the Bloch sphere, i.e., for a fixed
angle $\delta=\pi-\Delta\delta$ or $\delta=0+\Delta\delta$ with
$\Delta\delta\ll 1$, Eq. (\ref{new1}) simplifies to
\begin{eqnarray}
  F'(\delta)  = \frac18\left[ 5+\sqrt{2}
  +2\cos\Delta\delta-(\sqrt{2}-1)\cos(2\Delta\delta)\right]
\nonumber \\
 = 1-\frac{3-2\sqrt{2}}{8}(\Delta\delta)^2+ {\cal O}(\Delta\delta)^4.\hspace{21mm}
 \label{new2}
\end{eqnarray}

In a teleportation process, measurement of Alice results in four
possible outcomes $m_{i}$ where $i=0,1,2,3$ with
$m_{0}=|00\rangle\langle00|$, $m_{1}=|01\rangle\langle01|$,
$m_{2}=|10\rangle\langle10|$, and $m_{3}=|11\rangle\langle11|$.
Then the state at Bob's side conditioned on Alice's measurement
can be written as $\hat{\rho}'(m_{i})$. In the standard
teleportation protocol with shared MES, upon receiving the
classical information $i$, Bob can make the appropriate unitary
operations on his qubit $\hat{\rho}'(m_{i})$ to obtain the
teleported state $\hat{\rho}_{{\rm out}}=\hat{\rho}_{{\rm in}}$.
We discuss how the measurement result affects this process in the
presence of noise.

Since the entanglement distribution channel is noisy, the state at
the output of teleportation process given in Eq. (\ref{cc1}) can
be rewritten as $\hat{\rho}_{{\rm out}}={\rm Tr}_{\rm
in,a}[\hat{U}_{{\rm tel}}\hat{\rho}_{{\rm
in}}\otimes\hat{\rho}^{s}_{{\rm ent}} \hat{U}^{\dagger}_{{\rm
tel}}]$ where $\hat{\rho}^{s}_{{\rm ent}}$ is the noisy entangled
state. We can say that the fidelity is a function of $\delta$,
$\gamma$ and the noise introduced into the system, and we can
represent it as $F(\delta,\gamma)\equiv F(|\psi_{{\rm
in}}\rangle)$. We observe that $F(\delta,\gamma)$ is independent
of $\gamma$, as denoted by $F(\delta)\equiv F(\delta,\gamma)$.

\subsection{Amplitude damping channel}

\subsubsection{Input Bell states $|\psi^\pm\rangle$}

For the ADC, in scenario 1, let us assume that $p_{b}=p_{a}=p$ and
Alice made a measurement, obtained the outcome $m_{1}$ and then
sent the classical information $k=1$ to Bob. The output density
operator conditioned on $m_{1}$ becomes
$\hat{\rho}'(m_{1})=N[q\hat{\rho}_{{\rm
in}}+2p\sin^{2}(\delta/2)|0\rangle\langle0|]$ with $N$ being the
renormalization constant defined as $N^{-1}=1-p\cos\delta$ and
$\hat{\rho}_{{\rm in}}$ is the density operator of the state to be
teleported. Bob cannot rotate this $\hat{\rho}'(m_{1})$ to the
desired state without the prior knowledge of $\delta$ and
$\gamma$. Since $|\psi_{{\rm in}}\rangle$ is supposed to be
unknown, standard teleportation protocol fails to reproduce the
desired state at Bob's side. This conclusion is valid for all
$m_{i}$. Interestingly, output state at Bob's side can be grouped
into two as $\chi_{0}=\{\hat{\rho}'(m_{0}),\hat{\rho}'(m_{2})\}$
and $\chi_{1}=\{\hat{\rho}'(m_{1}),\hat{\rho}'(m_{3})\}$. Although
the output states in one of these groups can be rotated into each
other by using a Z-gate or first X then Z-gate, states belonging
to different groups cannot be rotated to each other. This problem
is caused by the ADC, which reduces the degree of entanglement and
introduces the additional terms
$2p\cos^{2}(\delta/2)|0\rangle\langle0|$ and
$2p\sin^{2}(\delta/2)|0\rangle\langle0|$, respectively, for
$\chi_{0}$ and $\chi_{1}$. We observed that for teleportation in
the presence of this ADC, if Alice's measurement yields $m_1$, Bob
does not need to do anything. For other measurement results
$m_0,m_2$ and $m_3$, Bob should apply $\hat \sigma_x$, $\hat
\sigma_y$ and $\hat \sigma_z$, respectively. In this way, he
rotates his qubit into the output state $\hat{\rho}_{\rm
out}(m_{k})=N_k[q\hat{\rho}_{{\rm
in}}+p(1+(-1)^{k}\cos\delta)|k\oplus 1\rangle\langle k\oplus 1|]$
with $N_k$ being the renormalization constant defined as
$N^{-1}_k=1+(-1)^k p\cos\delta$ and $\oplus$ stands for addition
modulo 2. Then the state-dependent fidelity becomes
\begin{eqnarray}\label{hytr1}
F_{m_{k}}(\delta)&=&1-\frac{p~(1+(-1)^{k}\cos\delta)^2}
{2(1+(-1)^{k}p\cos\delta)},
\end{eqnarray}
where $k=0,1,2,3$. When $p\rightarrow 1$, the limiting values are
calculated as $F_{m_{1,3}}(\delta)=\cos^2(\delta/2)$ and
$F_{m_{0,2}}(\delta)= \sin^2(\delta/2)$.

For $p\leq 1/11$ and $p< 1/5$, all states can be teleported,
respectively, with $F>5/6$ and $F>2/3$, independent of Alice's
measurement result. For the equatorial qubits $\delta=\pi/2$, we
find that as far as $p<1-\sqrt{2}/2$, teleportation fidelity will
surpass that of the PCCM regardless of the measurement outcome. On
the other hand, if the qubits are chosen at the neighborhood of
$|1\rangle$, then for even $k$ the channel damping rate should be
bounded as $0\leq p\leq 0.162$.

Although state-dependent teleportation fidelity is mainly
determined by Alice's measurement result, the average fidelity
calculated using Eq. (\ref{a1}) is the same for all measurement
results and given as
\begin{eqnarray}\label{hytr1ad}
\nonumber\\
F&=&\frac{1}{4 p^{2}}\left(2p+q^2\ln\frac{q}{1+p}\right),
\end{eqnarray}
which takes the minimum and maximum values of $1/2$ and $1$ for
$p=1$ and $p=0$, respectively.

For the second scenario, the output density operator elements
$\rho^{(jl)}_{\rm out}(m_{k})$ are found in terms of the input
density operator elements $\rho^{(jl)}_{\rm in}$ as follows:
$\rho^{(00)}_{\rm out}(m_{k})=q\rho^{(00)}_{\rm
in}+(1-(-1)^k)p/2$, $\rho^{(01)}_{\rm
out}(m_{k})=\sqrt{q}\rho^{(01)}_{\rm in} $, $\rho^{(10)}_{\rm
out}(m_{k})=\sqrt{q}\rho^{(10)}_{\rm in} $ and $\rho^{(11)}_{\rm
out}(m_{k})=q\rho^{(11)}_{\rm in}+(1+(-1)^k)p/2$. When Alice
measures $m_{k}$ and applies the appropriate unitary
transformation to get the highest fidelity for the process, i.e.,
when Bob receives the information that $k=3$ or $k=1$ for
$|\psi^+\rangle$ and $|\psi^-\rangle$ respectively, he can use a
Z-gate to rotate the state on his side to obtain the above state.
Then state-dependent fidelity can be found as
\begin{eqnarray}\label{hytr2}
F_{m_{k}}(\delta)&=&1-\frac12[p(1+(-1)^k\cos{\delta})-x],
\end{eqnarray}
where $x=(\sqrt{q}-q)\sin^2\delta$. In the limit of $p\rightarrow
1$, we get the same functions as those obtained from Eq.
(\ref{hytr1}).

It is clearly seen from above results that the range of qubits
that can be teleported correctly depends not only on the strength
of the ADC but also on the measurement result of Alice.

\begin{figure}[]
\hspace*{0mm} \epsfxsize=4.2cm
\epsfbox{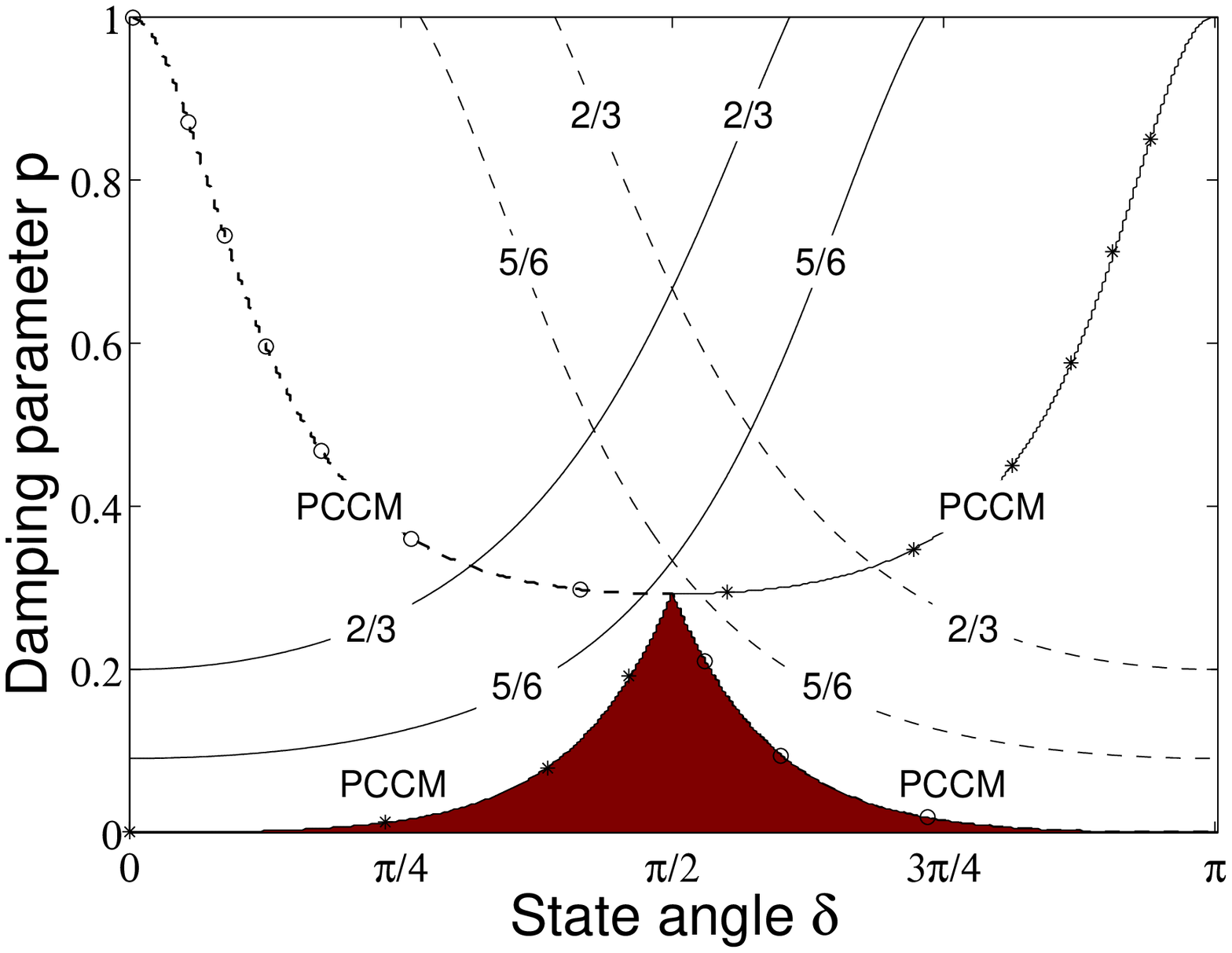}\hspace*{0mm}\epsfxsize=4.2cm
\epsfbox{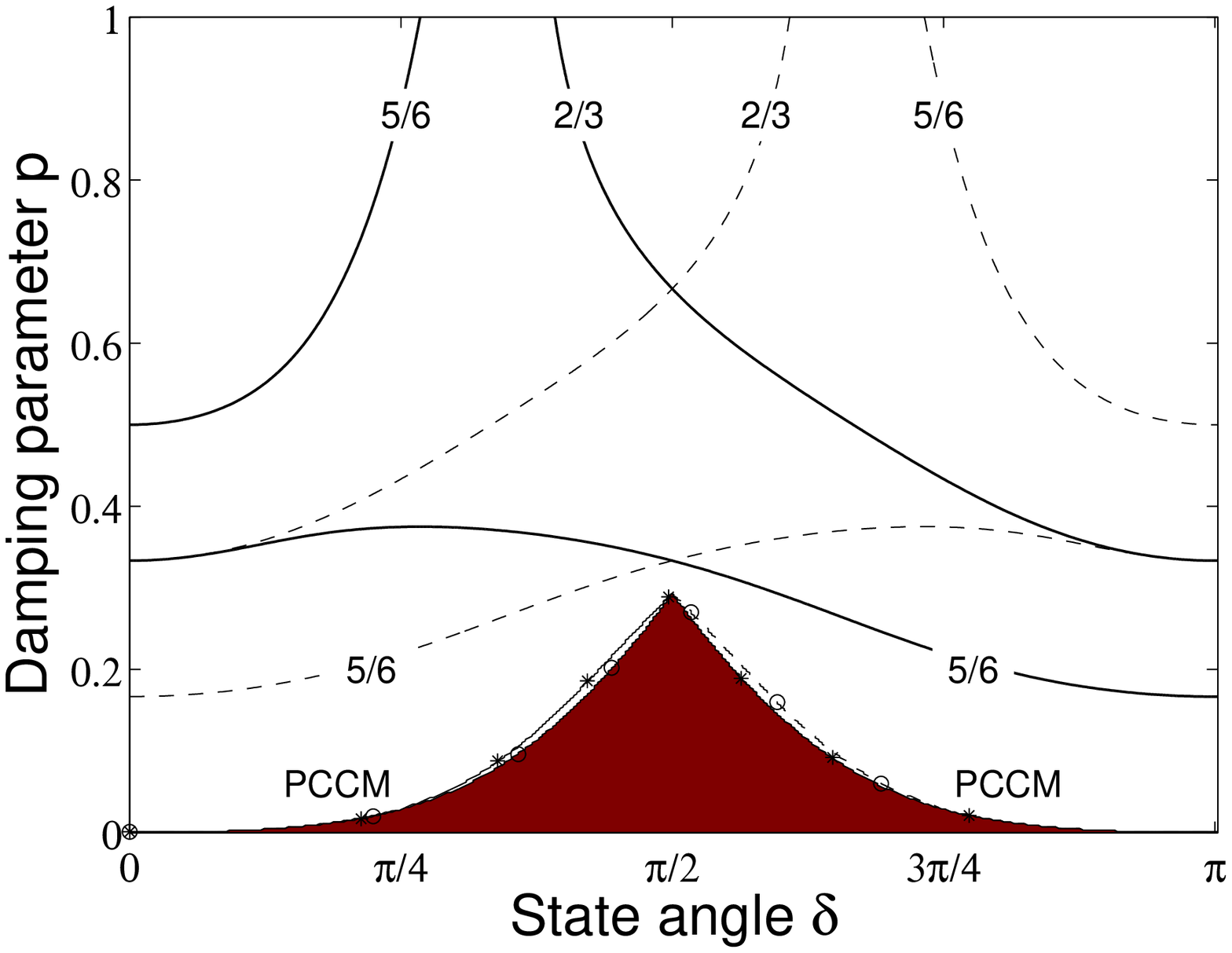}\vspace{0mm} \caption[]{Optimal state-dependent
fidelity in the presence of the ADC when the initial MES is
$|\psi^\pm\rangle$ (left) and $|\phi^\pm\rangle$ (right) and both
qubits are affected by damping. Contours correspond to $F=2/3$,
$F=5/6$ and the optimal PCCM fidelities when Alice's measurement
result is $|01\rangle\langle01|$ or $|11\rangle\langle11|$, solid
curves, and when Alice's measurement result is
$|10\rangle\langle10|$ or $|00\rangle\langle00|$, dotted
curves.}\label{fig3}
\end{figure}

\begin{figure}[]
\hspace*{-3mm} \epsfxsize=5cm
\epsfbox{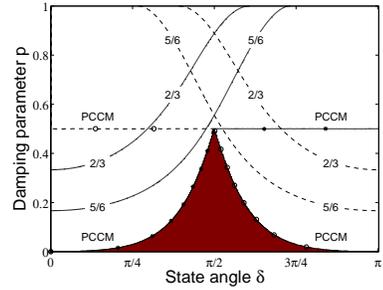}\hspace*{0mm}\epsfxsize=4cm
\vspace{3mm}\caption[]{Same as in Fig.~\ref{fig3} (left) for
$|\psi^\pm\rangle$ but for the case when only one of the qubits is
affected by the ADC. For $|\phi^\pm\rangle$, the meaning of curves
is reversed.}\label{fig4}
\end{figure}

Results imply that some of the states can be teleported with much
better fidelity than others depending on $m_{k}$. This enables
Alice and Bob, in a communication protocol, to decide to choose
their qubits randomly from a range of states with higher fidelity
when a certain measurement result, say $\{m_{0},m_{2}\}$, is
obtained. As seen in Figs. \ref{fig3} and \ref{fig4}, some states
give better fidelity than others depending on $m_k$. In the figure
we have shaded regions where all the states can be teleported with
$F>5/6$ regardless of Alice's outcome. Note that the states with
$\delta=\pi/2\mp\Delta$ can tolerate much higher damping rates
than the ones located around the poles of the Bloch sphere. It is
also seen that in Scenario 1 it is advantageous to rotate the
initial entangled state into $|\phi\rangle$ because it is more
immune to the ADC and therefore provides a larger parameter space
for $F>5/6$ teleportation.

Let us assume that qubits chosen from a range defined by $\Lambda$
have higher fidelity when Alice measures $\{m_{0},m_{2}\}$, on the
other hand qubits chosen from $\lambda$ have higher fidelity when
Alice measures $\{m_{1},m_{3}\}$. Then in a teleportation
protocol, Alice first mixes her state chosen from $\Lambda$ with
her part of the entangled state and makes a measurement, when she
obtains $\{m_{0},m_{2}\}$, she sends the other qubit of entangled
state to Bob together with the classical information, then Bob
applies unitary transformation to get the desired state. When she
gets $\{m_{1},m_{3}\}$, either she sends nothing or a dummy state.
In this way, they can increase the fidelity of the process. If
they decide to abort the protocol whenever Alice measures
$\{m_{1},m_{3}\}$ then the efficiency of the process is low.

If Alice and Bob decide to keep all measurement results then the
fidelity of the process can be written as
\begin{eqnarray}\label{pro1}
F_{{\rm}}(\delta)&=&\sum_{k=0}^{3}p_{m_{k}}F_{m_{k}}=q-\frac{x}{2}
\end{eqnarray}
where $x$ is defined as in Eq. (\ref{hytr2}), $F_{m_{k}}$ is the
fidelity of the output state to the teleported state when Alice
obtains $m_{k}$, and $p_{m_{k}}$ is the probability of obtaining
this result. Moreover, if they do that for any $(\delta,\gamma)$,
they end up with $F=2/3+(2\sqrt{q}-p)/6$. From Eq. (\ref{hytr2}),
it can be seen that for a fixed $p$ of the channel, if the state
to be teleported is chosen such that $\delta<\pi/2$, then the set
$\{m_{1},m_{3}\}$ gives higher teleportation fidelity for that
state; otherwise, the set $\{m_{0},m_{2}\}$ yields higher
fidelity. Let us assume that Alice randomly chooses a state to be
teleported from the upper hemisphere of the Bloch sphere,
$(\delta<\pi/2)$, therefore, their preferred measurement set is
$\{m_{1},m_{3}\}$, which occurs with a probability of $1/2$. When
the measurement result is $\{m_{0},m_{2}\}$, she sends nothing
according to the protocol described above. In this way, the
fidelity of the process increases to $F_{{\rm}}(\delta)=
F_{m_{1}}(\delta)=F_{m_{3}}(\delta)$ and the average fidelity
becomes $F=2/3+(4\sqrt{q}+p)/12$.

In the same way, if entangled state is distributed by a third
party and both qubits undergo damping, Alice proceeds as explained
above. If Alice and Bob decide to keep all measurement results
then the fidelity of the process becomes
\begin{eqnarray}\label{pro1a} F_{{\rm
}}(\delta)&=&\sum_{k=0}^{3}p_{m_{k}}F_{m_{k}}
=\frac{2-p(1+\cos^2\delta)} {2(1-p^2\cos^2\delta)},
\end{eqnarray}
where $F_{m_{k}}$ is the fidelity of the output state to the
teleported state when Alice obtains $m_{k}$, and $p_{m_{k}}$ is
the probability of obtaining this result. Moreover, if they do
that for any $(\delta,\gamma)$, they end up with
$F=\frac{1}{4p^2}[2p+q^2\ln(\frac{q}{1+p})]$. From Eq.
(\ref{hytr1}), it can be seen that for a fixed $p$ of the channel,
if the state to be teleported is chosen such that $\delta>\pi/2$,
then the set $\{m_{0},m_{2}\}$ gives higher fidelity; otherwise,
the set $\{m_{1},m_{3}\}$ does. Let us assume that Alice randomly
chooses the state to be teleported from the lower hemisphere of
the Bloch sphere, $(\delta>\pi/2)$, therefore, their preferred
measurement set is $\{m_{0},m_{2}\}$.

\subsubsection{Input states $|\phi^\pm\rangle$}
The output density operators for Alice's outcomes $m_{k=0,1,2,3}$
can be written as
\begin{eqnarray}\nonumber
\widehat{\rho }_{out}\left( m_{k}\right)&=&N[ q\widehat{%
\rho }_{in}+p\left( 1+(-1)^{k}p\cos \delta \right) \hat \sigma
_{x}^{k}\left\vert
0\right\rangle \left\langle 0\right\vert \hat \sigma _{x}^{k} \\
&& +(-1)^{k}pq\cos\delta \hat \sigma _{x}^{k}\left\vert
1\right\rangle \left\langle 1\right\vert \hat \sigma _{x}^{k}]
\end{eqnarray}
with $N^{-1}=1+(-1)^{k}p\cos \delta $ from which the state
dependent fidelity is derived as
\begin{equation}
F_{m_{k}}\left( \delta \right) =1-\frac{p\left[ 3-2p-(2p-1)\cos (2\delta )%
\right] }{4\left( 1+(-1)^{k}p\cos \delta \right) }
\end{equation}
with the limiting values of $F_{m_{0,2}}(\delta )=\cos ^{2}\left(
\delta /2\right) $ and $F_{m_{1,3}}(\delta )=\sin ^{2}\left(
\delta /2\right) $ for $p$ approaching $1.$
It is easy to see that as $p$ approaches $0$,  $%
F_{m_{k}}\left( \delta \right) \rightarrow 1.$ For $p<1/6$, all
states can be teleported with $F>5/6$ regardless of the outcome.
Average values of teleportation fidelity for these two cases are
the same as given in Eq. (\ref{hytr1ad}).

For the second scenario, contrary to the first scenario, the
output state that Bob gets after the proper application of the
quantum gates, and its fidelity to the desired state is the same
as that of the case when initial MES is $|\psi\rangle$. When only
one of the qubits of the MES goes through the ADC, distributing
either $|\psi\rangle$ or $|\phi\rangle$ does not give any
advantage to the parties.

\subsection{Phase damping channel}

\begin{figure}[h]
\hspace*{0mm} \epsfxsize=6cm
\epsfbox{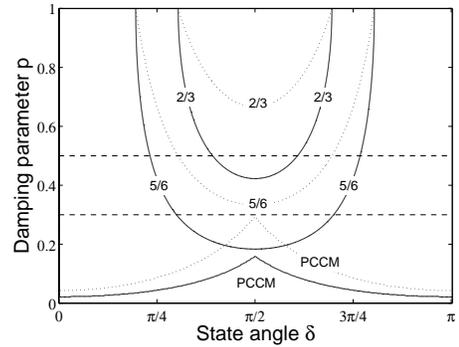}\vspace{10mm}\caption[]{State-dependent fidelity
in the presence of the PDC when the initial MES is any of the Bell
states and only one of the qubits (dashed curves) and both qubits
(solid curves) are affected by damping. Contours show $F=2/3$,
$F=5/6$ and the optimal PCCM fidelities. Horizontal dashed lines
correspond to the values of damping rate for $f_{{\rm ent}}=3/4$
for scenarios 1 (lower) and 2 (upper). }\label{fig5}
\end{figure}

When the channel is the PDC, given by (\ref{N05a}), only the
off-diagonal elements are affected by the damping. The fidelity of
the teleportation process when the initial MES is subjected to the
PDC is independent of Alice's measurement result, because,
contrary to the ADC case, Bob can use X and Z gates to rotate all
the possible outcomes to each other and to the state with the
highest fidelity to the input one. The elements of the density
matrix can be written as $\rho^{(00)}_{\rm out}=\rho^{(00)}_{\rm
in}$, $\rho^{(01)}_{\rm out}=q^{2}\rho^{(01)}_{\rm in}$,
$\rho^{(10)}_{\rm out}=q^{2}\rho^{(10)}_{\rm in}$ and
$\rho^{(11)}_{\rm out}=\rho^{(11)}_{\rm in}$ for the first
scenario. In case of the second scenario, the elements of the
density matrix are the same as above with $q^{2}$ replaced by $q$.
Then the fidelity of the teleportation process for scenario 1 can
be written as
\begin{eqnarray}\label{hytr3}
F(\delta)&=&1-\frac{1}{2}p(2-p)\sin^{2}\delta,\nonumber\\
F&=&1-\frac{1}{3}p(2-p),
\end{eqnarray}
and for scenario 2 as
\begin{eqnarray}\label{hytr4}
F(\delta)&=&1-\frac{1}{2}p\sin^{2}\delta,\nonumber\\
F&=&1-\frac{1}{3}p.
\end{eqnarray}
The effect of the PDC on the teleportation fidelity is the same
for both initial MES $|\phi\rangle$ and $|\psi\rangle$. The qubit
states with $\delta=0$ and $\delta=\pi$ which are located at the
poles of the Bloch sphere are always teleported with $F=1$.
Because these states correspond to $|0\rangle$ and $|1\rangle$,
which do not carry relative phase information and hence are not
affected by the PDC. Indeed, these results show that if Alice
chooses the states to be teleported around the poles then they can
have a better teleportation fidelity (see Fig.~\ref{fig5}). On the
other hand, states with $\delta=\pi/2$, which correspond to all
the states lying on the equator of the Bloch sphere, are the most
affected states.

If Eve does not have the information on the region from which the
qubits are chosen, then the best she can do is to use the optimal
universal quantum cloning machine of Bu\v{z}ek {\em et al. }
\cite{Buzek}. Then we find that any qubit state satisfying
$\sin^{2}\delta<1/3p(2-p)$ and $\sin^{2}\delta<1/3p$, respectively
for the first and second scenarios, can be teleported in the
presence of PDC  with higher fidelity than that of the cloning
machine of the eavesdropper. It is apparent that if there is no
eavesdropper and that parties just want to beat the classical
limit, the range of qubits at a fixed $p$ is much larger.

Now, let us assume that the states to be teleported are chosen
with fixed $\delta$ but varying $\gamma$, and the information on
$\delta$ may be leaked to an eavesdropper. Since the eavesdropper
may use the optimal PCCM, to speak about a secure teleportation
its fidelity should exceed the PCCM fidelity given in Eq.
(\ref{new2}). Comparing Eq. (\ref{new2}) with state dependent
teleportation fidelities for PDC given in Eqs. \ref{hytr3} and
\ref{hytr4}, we find $\cos\delta<-1+1/x'$ where
$x'=\sqrt{2}-1+2p(2-p)$ and $\cos\delta>1-1/x''$ where
$x''=\sqrt{2}-1+2p$, respectively for scenarios 1 and 2. If
$\delta$ is chosen in the neighborhood of $|1\rangle$, the damping
rate of the channel should satisfy $0\leq p<(3-2\sqrt{2})/4$ and
$0\leq p<1-(\sqrt{1+2\sqrt{2}})/2$, respectively for the first and
second scenarios. On the other hand, if the states to be
teleported are chosen from the equatorial qubit states, the
damping rate of the channel should satisfy $p<1-1/2^{1/4}$ and
$p<1-1/\sqrt{2}$, respectively, for the first and second
scenarios. These requirements are obviously stricter than those
for the universal CM. We see in Fig. \ref{fig5} that while for the
universal cloning machine the constraint on $p$ relaxes as we
approach to the poles of the Bloch sphere, for the PCCM it becomes
tighter. This is because as we approach the poles, the fidelity of
the clones from the PCCM gets closer to one requiring a PDC with
damping rates approaching to zero.

\subsection{Depolarizing channel}

When the channel is the DC, the elements of the density matrix can
be written as $\rho^{(00)}_{\rm out}=\chi\rho^{(00)}_{\rm
in}+\mu\chi$, $\rho^{(01)}_{\rm out}=\xi\rho^{(01)}_{\rm in}$,
$\rho^{(10)}_{\rm out}=\xi\rho^{(10)}_{\rm in}$ and
$\rho^{(11)}_{\rm out}=\chi\rho^{(11)}_{\rm in}+\mu\chi$ for the
first scenario, where we have used $\mu=(1-q)/2$ and $\chi=(1+q)$
and $\xi=q^{2}$. In case of the second scenario, the elements of
the density matrix are the same as above with $\chi=1$ and
 $\xi=q$. Then the fidelity of the teleportation process for the
first and second scenarios can be written, respectively, as
\begin{eqnarray}\label{hytr5}
F(\delta)=F=\frac12+\frac12 q^2,
\end{eqnarray}
\begin{eqnarray}\label{hytr6}
F(\delta)=F=\frac12+\frac12 q
\end{eqnarray}
from which we see that fidelity is independent of $|\psi_{{\rm
in}}\rangle$. For DC too, contrary (similar) to the ADC (PDC), Bob
can use quantum gates to rotate all possible outcomes to each
other. Therefore, the fidelity is independent of the input state,
of Alice's measurement result, and of the initially distributed
MES The parties in the protocol can choose their qubits from the
whole Bloch sphere and an eavesdropper may use a universal quantum
cloning machine, in that case the damping rates of the channels
should satisfy $p<1-\sqrt{6}/3$ and $p<1/3$ to surpass the
no-cloning limit. In case of an eavesdropper with the PCCM, the
relation between the qubits states that can be teleported securely
and the damping rate of the channel becomes
$\cos\delta<-(1+\sqrt{2})\left(-1+[1+4(\sqrt{2}-1)(xx-1)]^{1/2}\right)/2$
and
$\cos\delta<\left(-1+[17-12\sqrt{2}+8p(\sqrt{2}-1)]^{1/2}\right)/(2\sqrt{2}-2)$
for the first and second scenarios.

\subsection{Direct transmission: Noisy state $+$ shared MES}

For Bob to whom Alice wants to teleport the unknown state
$|\psi_{{\rm in}}\rangle$, it is difficult to distinguish whether
the $|\psi_{{\rm in}}\rangle$ is a noisy state or the quantum
channel is responsible for the noise. The state to be teleported
might be subjected to noise, loose its coherence and becomes a
mixed state before it is teleported.

Let us assume that Alice and Bob share a MES, which they have
obtained using entanglement distillation and purification
protocols. In this section, we assume that the qubit is influenced
by the ADC, PDC and DC, and discuss the outcome of the
teleportation process. We assume that only the qubit to be
teleported is subjected to noise and the shared entangled state is
any of the Bell states. Indeed, this is similar to direct
transmission scheme where the original state $|\psi_{{\rm
in}}\rangle$ is sent directly to Bob through noisy channel.

\begin{figure}[h]
\hspace*{-3mm} \epsfxsize=6cm
\epsfbox{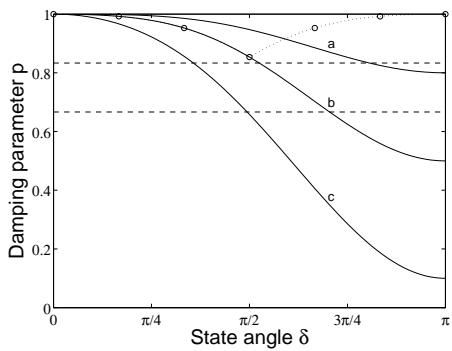}\hspace*{0mm}\epsfxsize=4cm \vspace{3mm}
\caption[]{State dependent teleportation fidelity when the qubit
to be teleported is subjected to the ADC, a, $p=0.2$, b, $p=0.5$,
and c, $p=0.9$. Horizontal dotted lines denote the limits between
classical and quantum operations (lower), and the secure quantum
teleportation (upper). The curve marked with circles corresponds
to the optimal PCCM fidelity.}\label{fig6}
\end{figure}

If $|\psi_{{\rm in}}\rangle$ is subjected only to the ADC, the
elements of the output density matrix become $\rho^{(00)}_{\rm
out}=\rho^{(00)}_{\rm in}+p~\rho^{(11)}_{\rm in}$,
$\rho^{(01)}_{\rm out}=\sqrt{q}~\rho^{(01)}_{\rm in}$,
$\rho^{(10)}_{\rm out}=\sqrt{q}~\rho^{(10)}_{\rm in}$ and
$\rho^{(11)}_{\rm out}=q\rho^{(11)}_{\rm in}$ where
$\rho^{(kl)}_{\rm in}$ are the elements of the density matrix of
$|\psi_{\rm in}\rangle$. Then fidelity is found as
\begin{equation}\label{bhy123}
F(\delta)=1-\frac{1}{2}[2p\sin^{2}(\delta/2)-(\sqrt{q}-q)\sin^{2}\delta].
\end{equation} Averaging this over all possible input states,
average fidelity is found as
\begin{equation}\label{bhy123a}
F=\frac{2}{3}+\frac{1}{6}(2\sqrt{q}-p),
\end{equation}
which is the same as for scenario 2, when the entangled state is
distributed through the ADC. We see that depending on the damping
parameter, the range of qubits that can be teleported with a
desired fidelity changes (see Fig.~\ref{fig6}). For example when
$p=0.8$, when only $|\psi_{\rm in}\rangle$ is subjected to noise,
the states with $\delta<0.5436\pi$ and $\delta<0.4021\pi$ can be
teleported, respectively, with $F>2/3$ and $F>5/6$. For the
$|\psi_{\rm in}\rangle$ damped case, all the states satisfying
$\delta<0.2677\pi$ can be teleported with $F>5/6$. In Fig.
\ref{fig6}, we have depicted the fidelity of the PCCM from which
we see that when $p=1/2$, the teleportation fidelity and the PCCM
fidelity are equal for the qubits $0\leq\delta\leq \pi/2$. In this
range of qubits, a secure teleportation is possible for damping
rates $p<1/2$. As $\delta$ approaches to $\pi$, the damping rate
$p$ approaches to zero to achieve secure teleportation.

When the qubit is subjected only to the PDC, the output density
operator becomes $\hat{\rho}_{{\rm out}} =q \hat{\rho}_{{\rm in}}
+p[\cos^{2}(\delta/2) |0\rangle
\langle0|+\sin^{2}(\delta/2)|1\rangle\langle 1|]$, resulting in a
fidelity $F=1-p\sin^{2}(\delta/2)$ from which average fidelity can
be written as $F=1-p/3$. Comparing these equations with Eq.
(\ref{hytr4}), it is seen that when the PDC affects only the qubit
to be teleported, the fidelity is the same as in scenario 2 when
the distributed entangled states undergo the PDC. We observe the
same similarity if only the qubit $|\psi_{\rm in}\rangle$ is
subjected to DC. In this case the fidelity expression is given as
in Eq. (\ref{hytr6}).

In the analysis of security of a damping particular channel, the
fidelities of optimal cloning machines were taken as a reference:
either (i) the optimal universal cloning machine if no a priori
information about a teleported state is given or (ii) the optimal
phase-covariant cloning machine if prior partial information about
the state is available. Clearly, a channel is secure if it
provides a better fidelity than the optimal cloning. This is the
lowest fidelity bound for security of any channel assuming that
Alice sends her qubit through a damping channel, while an
eavesdropper copies qubit at Alice's site and does not send it (or
sends it through a perfect channel). Otherwise, the action of the
channel will restrict the quality of the cloning consistent with
the channel and, thus, less demanding security conditions can be
given.

\begin{figure}[]
\hspace*{0mm} \epsfxsize=4.4cm
\epsfbox{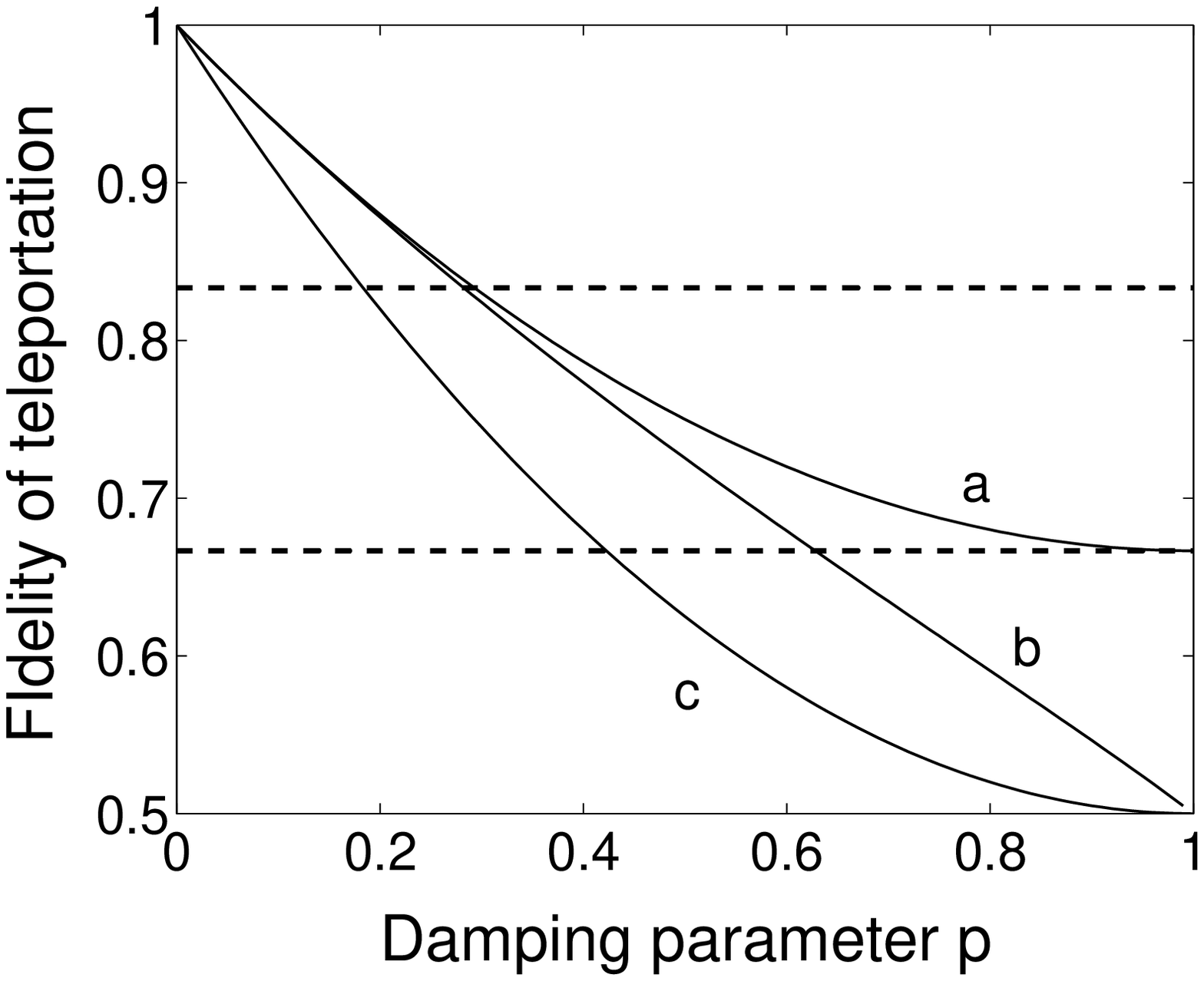}\hspace*{-3mm}\epsfxsize=4.4cm
\epsfbox{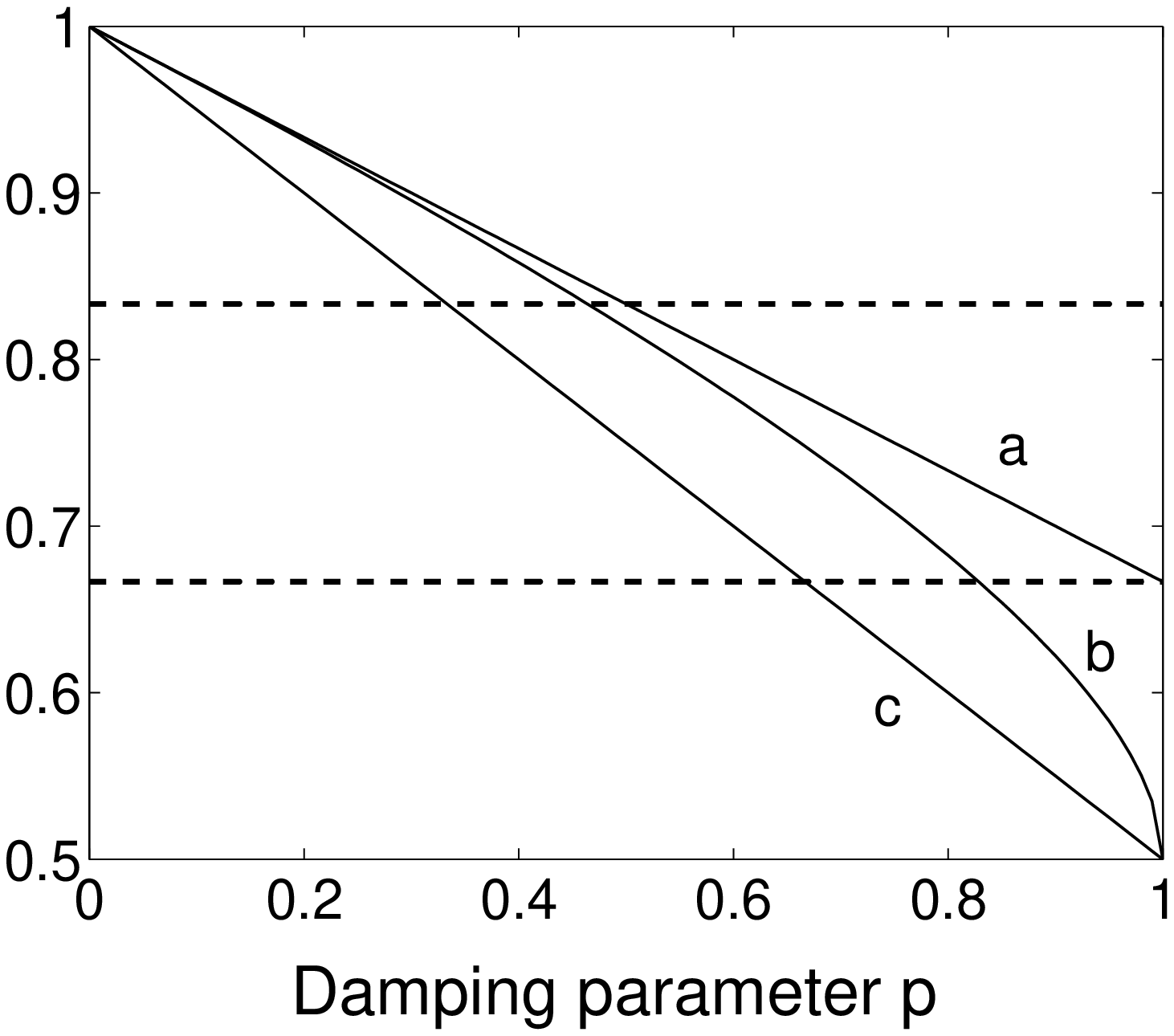}\vspace{0mm} \caption[]{Average fidelity with
noisy channels: a, PDC, b, ADC, and c, DC for scenarios 1 (left)
and 2 (right). Horizontal dashed lines denote the limits between
classical and quantum operations (lower), and the secure quantum
teleportation (upper). }\label{fig7}
\end{figure}

\section{CONCLUSION}

We have examined the problem of teleportation fidelity in the
presence of various types of noise during the entanglement
distribution of the teleportation process. Using the fully
entangled fraction and concurrence, we derived the bounds on the
damping parameters of channels so that the average fidelity (i)
exceeds the classical limit, and (ii) satisfy the security
condition for teleportation. Moreover, we derived the range of
states that can be teleported accurately with a desired fidelity
value and studied how this range is affected by noise. For the
security condition, we considered eavesdroppers with universal and
phase-covariant cloning machines where the first eavesdropper has
no information on the qubit to be teleported but in the latter
he/she knows the $\delta$ but not the relative phase $\gamma$.

For the ADC, although the bounds on $p$ for one-qubit affected
case are the same for both $|\psi^{\pm}\rangle$ and
$|\phi^{\pm}\rangle$ as the source entangled state, for the
two-qubit affected case we find that the bounds are different and
much tighter for $|\psi^{\pm}\rangle$. This implies that if one is
given $|\psi^{\pm}\rangle$, instead of distributing this state
directly, it is better to first locally convert to
$|\phi^{\pm}\rangle$ and distribute it. In that case the effect of
damping is less pronounced. We observe that only for the ADC these
bounds change with the initial MES to be distributed. We have
found that contrary to the case of the ADC, in the presence of the
PDC and DC, two-qubit affected case cannot be made to have higher
entangled fraction than the one-qubit affected case. Hence, the
average fidelity cannot be increased by subjecting one of the
qubits to controlled dissipation. As seen in Fig.~\ref{fig7},
average fidelity is dependent on the type and strength of damping
in the channel. For the PDC, fidelity is always larger than $2/3$
if $p\neq 1$, on the other hand for the ADC and DC average
fidelity decreases below $2/3$ down to $1/2$ depending on the
damping rate.

We have discussed the direct transmission case, too. We observe
that the results obtained for direct transmission and
teleportation with one-qubit affected entanglement distribution
case (scenario two), are the same in the cases of the DC and PDC.
However, discrepancies are seen for the case of the ADC. Average
fidelity for scenario 2 is more immune to damping than the direct
transmission.

This study shows that information on the noise affecting the
teleportation process during the phases of entanglement
distribution and the qubit preparation can be helpful in
increasing the fidelity. Moreover, it is important to note that if
the source of noise in the process is not known then all should be
attributed to an eavesdropper. Thus, the criterion on the
teleportation fidelity should be re-formulated taking into account
the set of states from where to be teleported state is chosen and
the optimal cloning machine for that set.

\begin{acknowledgments}
We thank Prof. Nobuyuki Imoto and Prof. Masato Koashi for their
useful comments. AM was supported by the Polish Ministry of
Science and Higher Education under Grant No. 1 P03B 064 28.
\end{acknowledgments}

\end{document}